\documentclass[12pt,preprint]{aastex}

\usepackage{emulateapj5}

\newcommand{\ml}{$M/L$}

\newcommand{\up}{$u$}
\newcommand{\gp}{$g$}
\newcommand{\rp}{$r$}
\newcommand{\ip}{$i$}
\newcommand{\zp}{$z$}

\newcommand{\hinv}{$h^{-1}\ $}

\newcommand{\mts}{$M_{260}$}
\newcommand{\mtsd}{$M_{260}^{dyn}$}
\newcommand{\mtwo}{$M_{200}$}
\newcommand{\lstar}{L$_{*}$}

\slugcomment{Accepted for publication in ApJ letters}

\shorttitle{Dynamical Confirmation of SDSS Weak Lensing Scaling Laws}
\shortauthors{}

\begin{document}


\title{Dynamical Confirmation of SDSS Weak Lensing Scaling Laws}


\author{
Timothy A. McKay\altaffilmark{1,10},
Erin Scott Sheldon\altaffilmark{1},
David Johnston\altaffilmark{2,10},
Eva K. Grebel\altaffilmark{3},
Francisco Prada\altaffilmark{3},
Hans-Walter Rix\altaffilmark{3},
Neta A. Bahcall\altaffilmark{4},
J. Brinkmann\altaffilmark{5},
Istv\'an Csabai\altaffilmark{6,7},
Masataka Fukugita\altaffilmark{8},
D.Q. Lamb\altaffilmark{2},
Donald G. York\altaffilmark{2,9}
}

\altaffiltext{1}{University of Michigan, Department of Physics, 500 East 
University, Ann Arbor, MI 48109}
\altaffiltext{2}{The University of Chicago, Department of Astronomy and 
Astrophysics, 5640 S. Ellis Ave., Chicago, IL 60637}
\altaffiltext{3}{Max-Planck-Institut f\"{u}r Astronomie, K\"{o}nigstuhl 17, 
D-69117 Heidelberg, Germany}
\altaffiltext{4}{Princeton University Observatory, Princeton, NJ 08544}
\altaffiltext{5}{Apache Point Observatory, P.O. Box 59, Sunspot, NM 88349-0059}
\altaffiltext{6}{Department of Physics and Astronomy, The Johns Hopkins 
University, 3701 San Martin Drive, Baltimore, MD 21218}
\altaffiltext{7}{Department of Physics of Complex Systems, E\"otv\"os 
University, P\'azm\'any P\'eter s\'et\'any 1}
\altaffiltext{8}{University of Tokyo, Institute for Cosmic Ray Reserach, 
Kashiwa, 2778582, Japan}
\altaffiltext{9}{The University of Chicago, Enrico Fermi Institute, 5640 S. 
Ellis Ave., Chicago, IL 60637}
\altaffiltext{10}{Center for Cosmological Physics, University of Chicago,
5640 S. Ellis Ave., Chicago, IL 60637}


\begin{abstract}
Galaxy masses can be estimated by a variety of methods; each applicable
in different circumstances, and each suffering from different systematic
uncertainties. Confirmation of results obtained by one technique
with analysis by another is particularly important. Recent SDSS weak lensing
measurements of the projected-mass correlation function reveal a linear
relation between galaxy luminosities and the depth of their dark matter
halos (measured on 260 \hinv kpc scales). In this work we use an entirely 
independent dynamical method to confirm these results. 
We begin by assembling a sample of 618 relatively isolated host galaxies,
surrounded by a total of 1225 substantially fainter satellites. We observe
the mean dynamical effect of these hosts on the motions of their satellites
by assembling velocity difference histograms. Dividing the sample by host
properties, we find significant variations in satellite velocity dispersion 
with host luminosity. We quantify these variations using a simple  
dynamical model, measuring \mtsd\, a dynamical mass within 260 \hinv kpc.
The appropriateness of this mass reconstruction is checked by  
conducting a similar analysis within an N-body simulation.
Comparison between the dynamical and lensing mass-to-light scalings shows
reasonable agreement, providing some quantitative confirmation for the 
lensing results.
\end{abstract}




\section{Introduction}

Mass clustered around galaxies
distorts the local space-time, bending light rays which pass near them
and inducing distortions 
in the images of more distant galaxies. Observation of these distortions
enables weak lensing measurements of the projected-mass correlation 
function (PMCF: \citet{bra96,fis00,wil01,hoe01}).
Recent measurements of the PMCF in fields drawn from the Las Campanas
Redshift Survey \citep{smi01} and the Sloan Digital Sky Survey 
(\citet{mck02}: hereafter M02) reveal an approximately linear 
correlation between the 
luminosity of L $>$ L$_{*}$ galaxies and the depth 
of the dark matter potential wells in
which they reside. This close connection is not particularly surprising
in a modern picture of galaxy formation \citep{sel00,guz01}, 
though it requires rather uniform 
integrated star formation efficiency across halos of a range of masses.

The weak lensing effects induced by galaxies are very small; peak 
distortions of 
background galaxies in the SDSS are only 0.5\%. In addition, quantitative
interpretation of the lensing results requires an accurate understanding
of lens geometries. 
%While lensing 
%methods provide many internal checks, and the source redshift
%distribution for these relatively shallow studies is thought to be well
%understood, it remains important to quantitatively confirm 
%these lensing scaling results by independent methods.
It is important to quantitatively confirm these lensing results by 
independent methods.

In this work we briefly examine galaxy mass-to-light scalings with an 
independent dynamical method. We extract a sample of `host' galaxies 
surrounded by
systems of `satellites' from data collected by the Sloan Digital Sky Survey.
Host galaxies typically have few satellites,
making it impractical to measure their masses individually.
By combining satellites from many hosts
of similar luminosity, we measure the mean velocity field around the hosts.
The method is analogous to the `stacking' of galaxies used in 
lensing measurements of the PMCF. It is closely 
related to the satellite galaxy analyses of \citet{zar93} and 
\citet{zar94}. 

\section{Data, and the Selection of Host and Satellite Galaxies}

Data for this study are drawn from Sloan Digital Sky 
Survey\footnote{www.sdss.org}: a combined imaging and spectroscopic survey of 
10$^4$ deg$^2$ in the North
Galactic Cap, and a smaller, deeper region in the South.
The imaging survey is carried 
out in drift-scan mode in five SDSS filters (\up, \gp, \rp, \ip, \zp)
to a limiting magnitude of r$<$22.5 \citep{fuk96, gun98}. 
The spectroscopic survey targets a `main' sample of 
galaxies with \rp$<$17.8 and a median redshift of z$\sim$0.1 
\citep{strauss01}. Velocity errors in the redshift survey are 
$\sim$30 km/s. For 
more details of the SDSS see \citet{yor00} and \citet{sto02}. 
For this study we extract from the SDSS main galaxy sample a catalog of 
102,922 galaxies with accurately measured redshifts in a range from 
z=0.01 to z=0.2.

Potential host galaxies must pass an isolation cut: they must be 
at least two times more luminous than any galaxy within a projected
distance of 2 \hinv Mpc and a velocity window of $\pm$1000 km/s. 
Around each such host, we select 
as possible satellites all galaxies at least
four times fainter than the host (1.5 magnitudes), within a projected
distance of 500 \hinv kpc, and within a velocity window of $\pm$1000 km/s.
Combining the requirement of faint satellites with the 
magnitude limit of the SDSS spectroscopic survey limits our host galaxies
to \rp $<$ 16.3. 

\begin{figure}
\epsscale{0.75}
\plotone{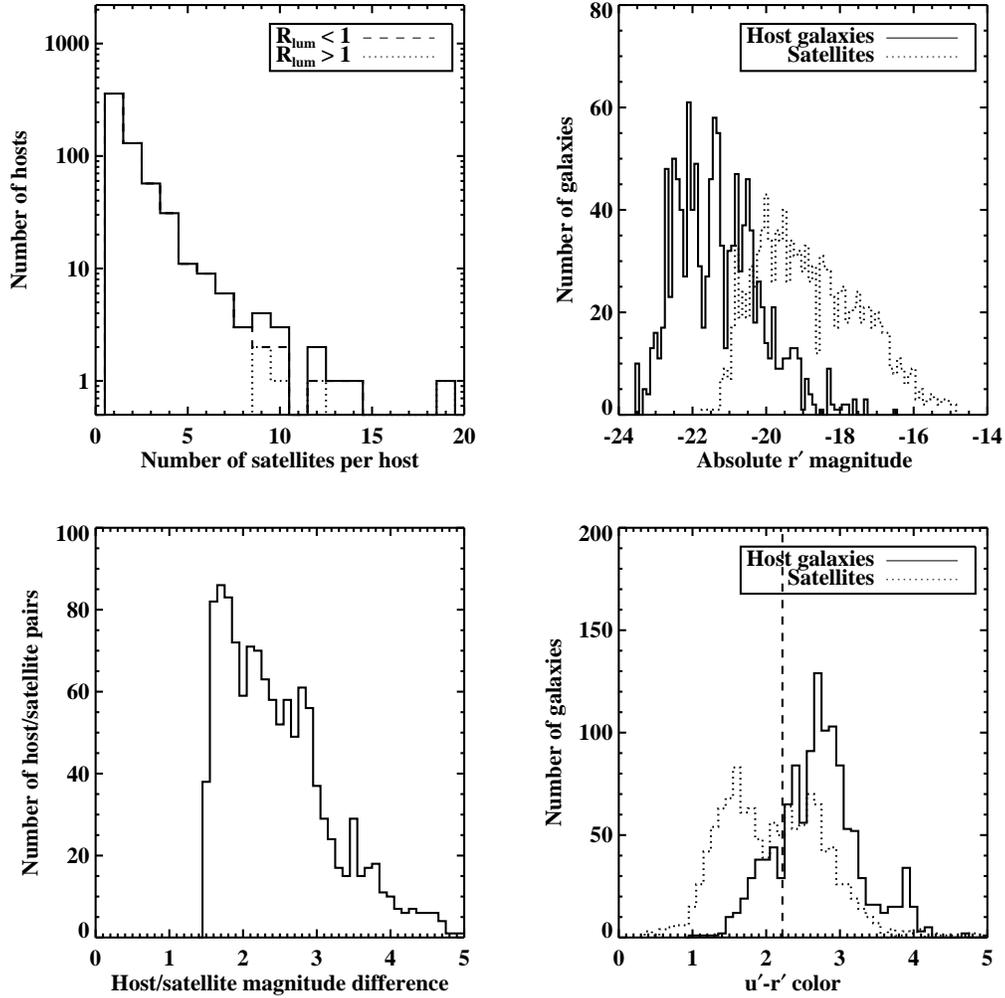}
\figcaption[f1.eps]{The four panels here provide an overview of 
the host and satellite populations selected by this method. The upper left
panel is a histogram of the number of satellites per host. The solid line is 
the full sample, the dashed (dotted) line is for those systems where the host 
galaxy is more (less) luminous than the sum of its satellites. The upper
right panel shows the absolute \rp\ magnitudes of the hosts and satellites 
in bins of 0.1 mag (using H$_{0}=100 h kms^{-1}Mpc^{-1}$ throughout).
The lower left panel shows the distribution of magnitude differences between
hosts and satellites. Finally, the lower right shows the distribution of
\up-\rp\ colors for hosts and satellites. The vertical dashed line in this
figure shows the approximate dividing line between early (redder) and 
late (bluer) type galaxies discussed in \citet{str01}. \label{basicinfo}}
\end{figure}

There are 13,737 galaxies with \rp$<$16.3 in the sample. Of these, 618 
are both isolated and have at least one faint satellite. Most of the 
remainder fail the isolation cut. The total number of satellites is
1225. While the mean number of satellites around each host is only two, 
there are host galaxies with as many as 19 satellites. An object
with this many satellites is clearly more a group with a dominant
central galaxy than a galaxy with a few satellites. We remove these 
by requiring that the combined luminosity of the satellites be less 
than the luminosity of the host. All but 5 of the host systems pass
this cut. Absolute magnitude distributions of host and 
satellite galaxies are shown in Figure \ref{basicinfo} . Many still fainter 
satellites exist; we are prevented from identifying them by 
the magnitude limit of the SDSS spectroscopic survey.

To reveal the mix of host galaxy types, we divide galaxies
into early (redder) and late (bluer) types using 
\up-\rp\ color as described
in \citet{str01}. The \up-\rp\ distribution for host 
and satellite galaxies is shown in the lower right panel of 
Figure \ref{basicinfo}. Overall the host galaxies
are 75\% early and 25\% late types. It is important to stress that this
division of early and late is luminosity dependent in a way which
is strongly color dependent. In \up, late types consitute 25\% of 
galaxies at all luminosities, while in \zp, late types are $>$ 50\% of
low luminosity galaxies, and $<$ 5\% of high luminosity galaxies.
This variation in the morphological mix with luminosity complicates 
comparison of these results to the M02 lensing results in the bluer
bands, where early and late type galaxies have substantially different
mass-to-light ratios.

\section{Dependence of Satellite Dynamics on Host Luminosity}

To examine the relationship between satellite dynamics and host galaxy
luminosity, we bin systems by host luminosity, then examine
the host-satellite velocity difference histograms for each luminosity bin.
All luminosities are based on SDSS Petrosian magnitudes, k-corrected in
the manner described in \citet{bla01}. An increase in the dispersion of 
satellite motions with host luminosity is clearly detected. 
%The lack of a correlation
%between host luminosity and satellite orbital motion in the data of
%\citet{zar94} may have been due to the relatively narrow range of
%host luminosities probed.

We quantify this variation 
by fitting the velocity difference histogram in each luminosity bin to the
sum of a Gaussian plus a constant, where the constant term represents the
contribution due to random interlopers not physically associated with the
host galaxy. The Gaussian widths fit to these 
distributions ($\sigma_r$) are direct
measures of the velocity dispersion characteristic of satellites for
host galaxies in each luminosity bin. While models of 
satellite motions \citep{whi92} suggest that the line of sight velocity
dispersion of satellites may be non-Gaussian, our measured profiles are well 
fit by these models. 
%In each case, there is a small but 
%statistically significant negative offset. These offsets are due to 
%selection effects outlined in \citet{zar92}. 
Errors quoted for these Gaussian
fits include the degeneracy between peak height and width. 

\section{Mass modeling}

To compare these results to the weak lensing results of M02, we
need to determine model masses for each luminosity bin. The M02 lensing
results quantify mass-to-light scalings by measuring \mts;
the mass inferred by fitting a singular isothermal sphere model to the 
observed density contrast within a radius of 260 \hinv kpc. We calculate 
here a very similar parameter, \mtsd. 

To derive this parameter we adopt a Jeans approach.
This approach is unlikely
to represent the situation in detail, as the dynamical times for satellites
at projected separations of 500 \hinv kpc approach the Hubble time. We
address this concern briefly by reference to N-body simulations in the 
following section. We begin with the 
relation \citep{bin87}:
\begin{equation}
\frac{GM(r)}{r} = -\overline{v^{2}_{r}} \left (
	\frac{\partial{\ln{\nu}}}{\partial{\ln{r}}} +
	\frac{\partial{\ln{\overline{v^{2}_{r}}}}}{\partial{\ln{r}}} +
	2 \beta \right )
\label{bt4-55}
\end{equation}
where r is the radius at which we measure the mass, 
$\overline{v^{2}_{r}}$ is the average radial velocity of satellites
squared, 
$\nu(r)$ is the
number density of satellites as a function of radius, and $\beta$ is 
the velocity anisotropy 
\begin{equation} 
\beta = 1.0 - \frac{\overline{v^{2}_{\theta}}}
	{\overline{v^{2}_{r}}}
\end{equation}
By measuring
the projected number density of satellites as a function of radius
we determine $\nu(r) \propto r^{-2.1}$. 
%We probe $\partial{\ln{\overline{v^{2}_{r}}}} / \partial{\ln{r}}$ by 
%measuring $\sigma_r$ in apertures from 100 to 500 \hinv kpc. The resulting
%$\sigma_r$ measurements are all consistent within 1 $\sigma$, implying
%that the second term in the parentheses of Equation \ref{bt4-55} is
%zero. 
Measurements of $\sigma_r$ in apertures from 100 to 500 \hinv kpc show no
significant variation, implying
that the second term in the parentheses of Equation \ref{bt4-55} is
zero. We assume that the velocity anisotropy $\beta$=0. 
If this is
incorrect, the masses determined here will be biased by a factor of 
$\sim$(1.0 - $\beta$). Finally, we assume that our measured $\sigma_r^2$,
the mean square line of site velocity of all satellites within a 
projected distance of 500 \hinv kpc of the host, is equal to 
$\overline{v^{2}_{r}}$. These are significant assumptions, which we test
below by simulation.

With these assumptions, our mass estimator reduces to 
the simple form:
\begin{equation}
M_{260}^{dyn} = \frac{2.1\times r\times \overline{v^{2}_{r}}}{G}
\label{m260_def}
\end{equation}
where r is the 260 \hinv kpc radius to which we integrate the mass, and 
$\overline{v^{2}_{r}}$ is the one dimensional RMS satellite velocity.

%It is likely
%that one or more of the assumptions made above is inappropriate. 
%To first order, this is likely to shift the mass scale. But it 
%is possible that one or more of the terms in Equation \ref{bt4-55} is
%dependent on host galaxy luminosity. These issues will be addressed further
%as SDSS satellite galaxy samples grow larger.

\section{Comparison to GIF simulations}

A variety of assumptions were made to arrive at Equation \ref{m260_def}. It 
is possible that one or more of these assumptions introduces systematic
error. In addition,
it is desirable to relate the mass esimator \mtsd\ to theoretically 
favored quantities like
\mtwo. As a first test, we conduct a parallel analysis within a simulated 
universe, extracting the observables used in our analysis from the 
simulation. By reconstructing masses in the simulation using \mtsd, and 
comparing these to \mtwo, it is possible to both test the validity of \mtsd 
and understand its relationship to \mtwo.

To conduct this test we turn to the GIF simulations of \citet{kau99}. GIF 
combines N-body simulations of the evolution of dark matter clustering
with a semianalytic galaxy modeling scheme, including gas cooling, 
star formation, feedback
from supernovae, and galaxy merging. The end result is a catalog
of galaxies with B, V, R, I, and K magnitudes and stellar masses. Peculiar 
velocities for each galaxy 
are determined from the N-body simulations. As a result, these
velocities have the full structure we might expect to find in the data. The 
most important limitation of the GIF simulations for this comparison is 
their relatively low mass resolution. They resolve only the 
most massive and luminous host galaxies.

To test our analysis of SDSS satellite motions, we conduct a parallel 
analysis within the GIF simulations. We define as isolated all galaxies 
which are at least two times brighter than any other galaxy within their 
halo. We then search these for cases with satellites at least four times 
fainter. We use the velocity differences between hosts and satellites to 
populate velocity difference histograms, and sort these by host luminosity, 
in the same manner used in the SDSS satellite study. As a first check of
Equation \ref{m260_def}, we use the simulations to check our assumptions 
about $\beta$ and the relationship between $\sigma_{r}^{2}$ and 
$\overline{v_{r}^{2}}$. We calculate the velocity anisotropy from
the simulation, and find $\beta = 0.06 \pm 0.03$. We also 
find the LOS velocity dispersion $\sigma_{r}^{2} = 1.03 \pm 
0.03 \times \overline{v_{r}^{2}}$. While
the GIF simulations are not a perfect match to our satellite sample, this
provides some support for the assumptions we made in deriving Equation 
\ref{m260_def}.

\begin{figure}
\plotone{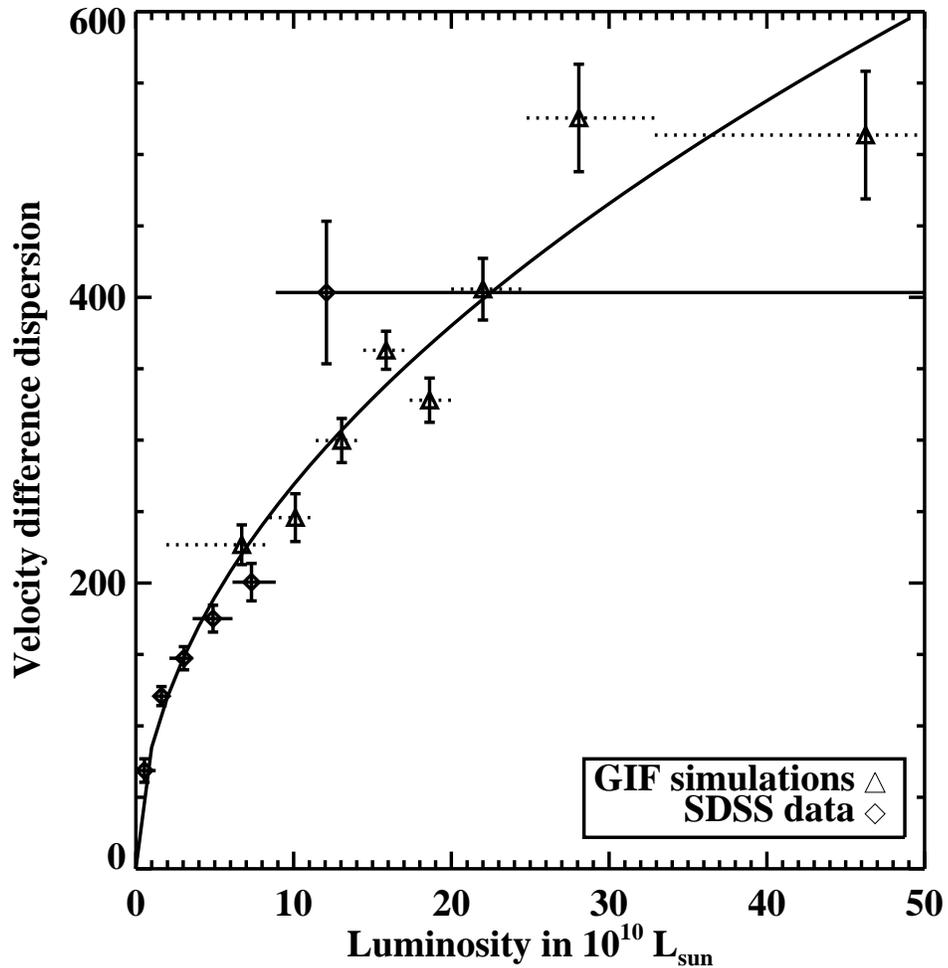}
\figcaption[f2.eps]{This figure compares measurements 
of satellite velocity dispersion vs. luminosity for real SDSS satellites 
(diamonds) and simulated GIF satellites (boxes). The two measures are well 
represented by a single relationship $\sigma_r \propto L^{0.5}$. Vertical 
error bars represent the uncertainties in determination of the satellite 
velocity dispersion. Horizontal error bars represent the range of host 
luminosities for each bin. The results from GIF simulations allow us to 
test the applicability of our mass estimator. The similarity of the simulations
and observations at this observable level give some confidence in the comparison.
\label{sdss_gif_comparison}}
\end{figure}

A more direct approach is to confirm that the data and simulations show a 
consistent relationship between the observables: $\sigma_{r}$ and luminosity.
This comparison of the 
SDSS and GIF $\sigma_{r}$ vs. luminosity measurements is shown in 
Figure \ref{sdss_gif_comparison}. The SDSS luminosity measurements are in 
\ip, and the GIF luminosities in Johnson I, but as both are converted to 
solar luminosities the comparison remains appropriate. Both the SDSS 
and GIF results are consistent with a simple model in which $\sigma_{r} 
\propto L^{0.5}$.

Since the GIF simulations provide values for the \mtwo\ associated with 
each halo, we can compare measures of \mtsd\ derived from galaxy 
velocities to \mtwo\ within the simulations. This comparison suggests 
that \mtsd $\approx 0.7\times$\mtwo. It is essential to note that this
relation is determined only for the most luminous galaxies, and over only a 
factor of four in luminosity. While this comparison is limited, it
provides some confidence that Equation \ref{m260_def} sensibly relates 
satellite
velocities to masses. A more detailed understanding awaits simulations with
higher mass resolution, in which more direct comparisons can be made.

\section{Mass-to-light scalings and comparisons to lensing results}

\begin{figure}
\epsscale{0.75}
\plotone{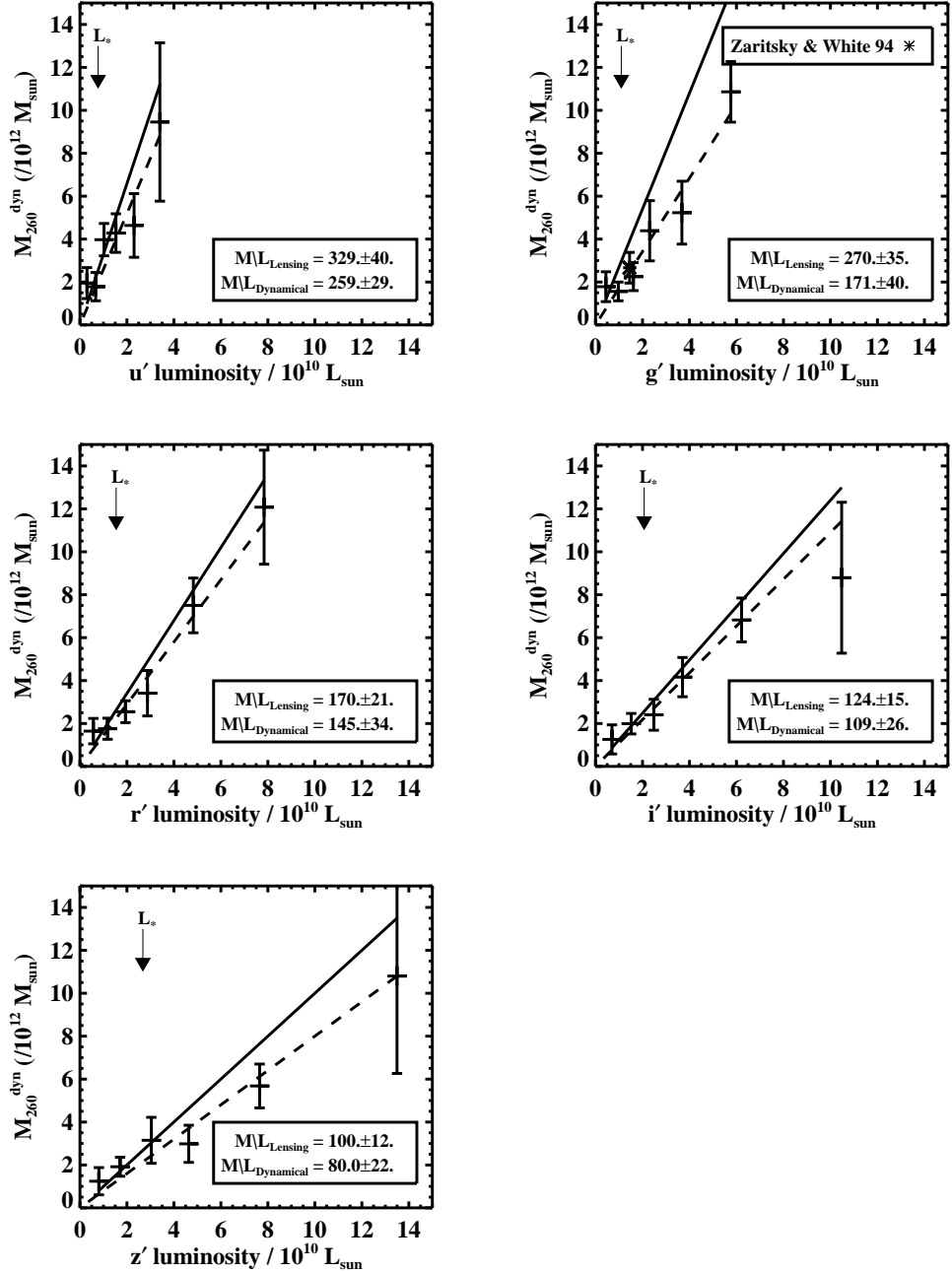}
\figcaption[f3.eps]{This figure shows the 
relationship between
\mtsd\ and luminosity in each of the five SDSS bands.
In each plot the data points are the \mtsd estimates for each luminosity
bin. In the \gp\ plot, an additional point has been added giving the 
approximate mass derived from satellite studies of \citet{zar94}, scaled
to a radius of 260 \hinv kpc assuming isothermality. The vertical arrow
in each plot marks the luminosity of an \lstar\ galaxy in each band.
The solid line in each plot represents the best fit lensing \ml\ for
from M02. The dashed lines represent the best fit constant \ml\ model from
these dynamical measurements. Note that all five figures are on 
the same scale. Most of the variation of mass with luminosity is seen 
for L $>$ \lstar\ galaxies.
\label{mass_comparison}}
\end{figure}

The measurements of \mtsd\ derived from satellite motions are shown in 
Figure \ref{mass_comparison}. The relationship between \mtsd\ and light 
in each passband is well fit by a single power law. As observed in M02, 
these relations are consistent with a power law index of one (constant 
\ml) in all bands except \up, where a flatter relation, with a best fit 
power law index $\sim$0.6 is observed. 

To compare these results to the M02 lensing results, we fix the power law 
index to one,  and fit the \mtsd\ vs. luminosity data in each band to 
obtain values for \ml. Values for these \ml\ ratios are given in
Figure \ref{mass_comparison}. The dynamical \ml\ values are consistent
with the lensing values at the 1$\sigma$ level in most bands. They differ 
most strongly in 
u and g, where M02 suggests results will be very sensitive to the
mix of host types. 

The masses of L$_{*}$ galaxy halos are proportional to 
light, and \ml\ 
values vary strongly with color. 
More detailed quantitative comparison of the lensing and dynamical 
results will only be possible when we can measure more identical samples of
hosts and lenses. This is difficult. The magnitude limit of the 
SDSS spectroscopy biases dynamical studies toward hosts at z$<$0.05, 
while lens geometries prefer lenses at z$\sim$0.15. Still, the \ml\ values 
derived by independent methods agree reasonably, especially in the redder 
bands, where differences between the host and lens samples are probably less 
important. 

\section{Conclusions}

Recent SDSS weak lensing results (M02) revealed a linear relation 
between galaxy luminosity and mass on halo scales. 
To quantitatively test this result,
we have measured the relationship between galaxy luminosity and satellite 
dynamics for a large sample of reasonably isolated host galaxies. We observe 
a highly significant increase in satellite velocity with host luminosity.

To make direct comparisons to the weak lensing results, we apply a 
simple mass estimator, \mtsd, which is closely analogous to the 
mass modeling used in M02. A first order test
of the validity of this model is made by computing it within the GIF 
simulations. The relationship 
seen here between \mtsd\ and luminosity matches reasonably
the relation seen by entirely independent weak lensing methods. This 
confirms the essential conclusions reached in M02. The luminous and dark 
components of L$_{*}$ galaxies are strongly coupled on $\sim$200 kpc 
scales. 

It is unknown whether the scaling relations observed
here for super-\lstar\ galaxies apply at lower luminosity. Perhaps the 
most important future extension of these studies will be to less 
luminous galaxies. Additional comparison of lensing and dynamical mass
estimates will require measurement of more closely related host and 
lens samples, and more detailed study of the extraction of mass from
the observable PMCF and satellite velocity structure.

\acknowledgments
The Sloan Digital Sky Survey (SDSS) is a joint project of The University 
of Chicago, Fermilab, the Institute for Advanced Study, the Japan 
Participation Group, The Johns Hopkins University, Los Alamos National
Laboratory, the Max-Planck-Institute 
for Astronomy (MPIA), the Max-Planck-Institute for Astrophysics (MPA), 
New Mexico State University, Princeton University, the United States Naval
Observatory, and the University of Washington. Apache Point Observatory, 
site of the SDSS telescopes, is operated by the Astrophysical Research 
Consortium (ARC). 

Funding for the project has been provided by the Alfred P. Sloan Foundation, 
the SDSS member institutions, the National Aeronautics and Space 
Administration, the National Science Foundation, the U.S. Department of 
Energy, the Japanese Monbukagakusho, and the Max Planck Society. The
SDSS Web site is http://www.sdss.org/. Timothy McKay and Erin Sheldon 
gratefully 
acknowledge support from NSF PECASE grant AST 9708232. We thank the GIF
team for making their simulations publically available, and acknowledge
Simon White and the anonymous referee for constructive suggestions.

\clearpage

%\begin{figure}
%\epsscale{0.75}
%\plotone{vdisp_lum_final.ps}
%\figcaption[vdisp_lum_final.ps]{This figure shows the velocity difference histogram
%for host galaxies of different \rp\ luminosities. The lowest luminosity sample 
%is on the lower right, the highest on the upper left. A strong increase in 
%satellite velocities with host \rp\ luminosity is observed. Similar increases
%are observed when the hosts are grouped by their luminosities in other bands.
%\label{vdisp_lum}}
%\end{figure}

%%%%%%%%%%%%%%%%%%%%%%%%%%%%%%%%%%%%%%%%%%%%%%%%%%%%%%%%%%%%%%%%%%%%%%%%%%%
% Table of results of power law fits to the dynamical mass in each passband
%%%%%%%%%%%%%%%%%%%%%%%%%%%%%%%%%%%%%%%%%%%%%%%%%%%%%%%%%%%%%%%%%%%%%%%%%%%%


\begin{thebibliography}{}
%\bibitem[Berlind and Weinberg(2001)]{ber01} Berlind, A., and Weinberg,
%	D.\ 2001, astro-ph/0109001
\bibitem[Binney and Tremaine(1987)]{bin87} Binney, J., and Tremaine, S.,
	1987, Galactic Dynamics (Princeton, NJ: Princeton University Press)
\bibitem[Blanton et al.(2001)]{bla01} Blanton, M.~R.~et al.\ 2001, \aj, 
	plot121, 2358
\bibitem[Brainerd, Blandford and Smail (1996)]{bra96} 
	Brainerd, T. G., Blandford, R. D. and Smail, I. 1996, \apj, 466, 623 
\bibitem[Eisenstein, et al.(2001)]{eis01} Eisenstein, D., et al.\ 2001,
	accepted for publication in \aj 
\bibitem[Fischer et al.(2000)]{fis00} Fischer, P.~et al.
	2000, \aj, 120, 1198 
\bibitem[Fukugita et al.(1996)]{fuk96} Fukugita, M., 
        Ichikawa, T., Gunn, J.~E., Doi, M., Shimasaku, K., \& 
	Schneider, D.~P.\ 1996, \aj, 111, 1748 
\bibitem[Gunn et al.(1998)]{gun98} Gunn, J.~E.~et al.\ 1998, 
	\aj, 116, 3040 
\bibitem[Guzik \& Seljak(2001)]{guz01} Guzik, J.~\& Seljak, 
	U.; 2001, \mnras, 321, 439 
\bibitem[Hoekstra, Yee, \& Gladders(2001)]{hoe01} Hoekstra, H., Yee, H., 
	and Gladders, M., 2001, submitted to ApJ, also astro-ph/0107413
\bibitem[Kauffman, et al.(1999)]{kau99} Kauffman, G., Kolberg, J., Diaferio,
	A., \& White, S.D.M., 1999, \mnras, 303, 188
\bibitem[McKay, et al.(2002)]{mck02} McKay, T., et al., 2002, submitted
	to \apj, also astro-ph/0108013
%\bibitem[Sheldon et al.(2001)]{she01} Sheldon, E.~S.~et al.\ 2001, \apj, 
%	554, 881
\bibitem[Seljak(2000)]{sel00} Seljak, U., 2000, \mnras 318, 203
\bibitem[Smith, Bernstein, Fischer, \& Jarvis(2001)]{smi01}
 	Smith, D.~R., Bernstein, G.~M., Fischer, P., \& Jarvis, M.\ 2001, 
	\apj, 551, 643
\bibitem[Stoughton et al.(2002)]{sto02} 
	Stoughton, C.~et al.\ 2002, \aj, 123, 485
\bibitem[Strateva, et al.(2001)]{str01} Strateva, I., et al.\ 2001,
	accepted for publication in \aj, also astro-ph/0107201
\bibitem[Strauss, et al.(2001)]{strauss01} Strauss, M., et al., in preparation
\bibitem[York et al.(2000)]{yor00} York, D.~G.~et al.\ 2000, 
	\aj, 120, 1579 
\bibitem[White \& Zaritsky(1992)]{whi92} White, S.~D.~M.~\& Zaritsky, 
	D.\ 1992, \apj, 394, 1
\bibitem[Wilson, Kaiser, Luppino, \& Cowie(2001)]{wil01} Wilson, G., 
	Kaiser, N., Luppino, G.~A., \& Cowie, L.~L.\ 2001, \apj, 555, 572
\bibitem[Zaritsky(1992)]{zar92} Zaritsky, D.\ 1992, \apj, 
        400, 74 
\bibitem[Zaritsky, Smith, Frenk, \& White(1993)]{zar93} Zaritsky, D., 
	Smith, R., Frenk, C., \& White, S.~D.~M.\ 1993, \apj, 405, 464 
\bibitem[Zaritsky \& White(1994)]{zar94} Zaritsky, D.~\& White, S.~D.~M.\ 
	1994, \apj, 435, 599 
\end{thebibliography}
\end{document}